\documentclass[cameraready]{Interspeech}
\usepackage{amsmath}
\usepackage{amssymb}
\usepackage{bm}
\usepackage{booktabs}
\usepackage{graphicx}
\usepackage{url}

\title{Zero-Shot Face-to-Speech Synthesis via Latent Space Adaptation of a Style-Diffusion TTS Model}

\author[affiliation={1,2}, orcid=0009-0008-1193-194X]{Carlos}{Muñoz-Romero}
\author[affiliation={3,1}, orcid=0000-0002-5531-8994]{Jose A.}{Gonzalez-Lopez}

\address{
    $^1$ Universitat Oberta de Catalunya (UOC), Spain \\
    $^2$ Monoceros Labs, Spain \\
    $^3$ Dpt.\ of Signal Theory, Telematics and Communications,  University of Granada,  Spain
}

\email{carlosmr@uoc.edu, joseangl@ugr.es}


\begin{document}

\maketitle

%==================================================================================
% Abstract (<= 1000 characters, ASCII, no citations).
%==================================================================================
\begin{abstract}
Zero-shot text-to-speech (TTS) clones a voice from a short audio prompt, but this reliance on reference audio is a barrier when only visual information is available, e.g. for historical figures or video-game characters. In this work, we propose a Face-to-Speech (F2S) framework that predicts a plausible voice from a static facial image. A lightweight Face Adapter, together with soft-tuning of the face encoder's upper blocks, aligns face-recognition features with the style space of a frozen StyleTTS 2 model, kept frozen during training. We evaluate on held-out identities from LRS3, a large-scale audiovisual corpus of English TED-talk videos. The synthesized speech is highly natural (UTMOS 3.7-4.0, matching or exceeding the 3.61 of ground truth), face-to-voice retrieval is consistently above chance, and the generated voice is consistent with the target speaker. Without any retraining, an English-trained adapter also produces fluent Spanish speech, indicating that the face-to-style mapping is largely language-agnostic.
\end{abstract}

\section{Introduction}

\label{sec:intro}
Recent advances in zero-shot text-to-speech (TTS) enable voice cloning from a few seconds of reference audio \cite{Tan21-Survey, Zhu25-ZipVoice, Zhou25-VoxCPM, Peng25-VibeVoice}. Systems such as Tacotron~2 \cite{Shen18-Tacotron2}, FastSpeech~2 \cite{Ren21-FastSpeech2} and HiFi-GAN \cite{Kong20-HiFiGAN}, along with recent large-corpus models (ZipVoice \cite{Zhu25-ZipVoice}, VoxCPM \cite{Zhou25-VoxCPM}, VibeVoice \cite{Peng25-VibeVoice}), clone an arbitrary voice from a short audio prompt. StyleTTS~2 \cite{Li23-STTS2}, in particular, models prosody as a latent style sampled by a diffusion model, reaching human-level naturalness. This paradigm is, however, fundamentally constrained when no vocal recording exists, such as generating voices for historical figures, fictional characters, or non-player characters (NPCs, i.e., computer-controlled characters) in video games.
Face-to-Speech (F2S) synthesis bypasses this by predicting vocal identity directly from facial attributes \cite{Lee24-FVTTS}, biologically grounded in the shared developmental link between facial bone and vocal-organ morphology (e.g.\ androgens jointly drive mandibular growth and vocal-fold thickening) \cite{Smith16-Matching, Mavica13-Matching}. Early systems predicted a speaker embedding from a face: Face2Speech \cite{Goto20-Face2Speech} drove a multi-speaker TTS this way, while FVTTS \cite{Lee24-FVTTS} generated a coherent voice end-to-end, reporting a Speaker Encoder Cosine Similarity (SECS, cosine between speaker embeddings of generated and reference audio) of 0.754 on LRS3 \cite{Afouras18-LRS3}. Diffusion-based F2S \cite{Lee23-ImaginaryVoice} preserves a face-derived identity, and later work decouples identity from prosody \cite{Kang23-FaceStyleSpeech} or emotion \cite{Ye25-EmotionalF2S} on corpora such as VoxCeleb2 \cite{Chung18-VoxCeleb2}.
Despite this potential, F2S faces a critical challenge: the weak, non-deterministic correlation between facial and vocal traits. Approaches trained with simple regression objectives tend to suffer from \emph{mode collapse} \cite{Salimans16-ImprovedGAN}, where generated voices converge toward generic, gender-averaged prototypes \cite{Goto20-Face2Speech}. At its core, F2S is an alignment between heterogeneous latent spaces: Deep CCA \cite{Andrew13-DCCA} captures high-variance shared attributes (e.g.\ gender) but lacks intra-class discrimination, and pair-based metric learning \cite{Hadsell06-DRLIM} is unstable against a fixed target. Extending F2S beyond English is further under-explored due to the scarcity of paired multilingual audiovisual data.
This paper addresses these limitations with a ``Freeze-Align'' F2S framework.\footnote{Links to the code repository and audio samples will be provided upon paper acceptance.} We keep a state-of-the-art acoustic generator (StyleTTS 2 \cite{Li23-STTS2}) frozen as a teacher and train a lightweight adapter that maps faces into its style space \cite{Maniparambil25-Frozen}---a teacher-side alignment validated for audio by Wav2CLIP \cite{Wu22-Wav2CLIP}---while also soft-tuning the upper blocks of the face encoder (the majority of its parameters). Our contributions are:

\begin{itemize}

\item A \textbf{Freeze-Align} F2S system: a Face Adapter over a \emph{frozen} StyleTTS 2 acoustic teacher (with soft-tuning of the face encoder) that synthesizes natural speech from a face without training the generator.

\item An \textbf{inference-time control} that decouples face-derived timbre from prosody and exposes an explicit, tunable identity-vs-naturalness trade-off.

\item A \textbf{hybrid contrastive loss} (Relational Knowledge Distillation \cite{Park19-RKD} plus demographic auxiliary heads) that structures the face-voice latent space against mode collapse.

\item A \textbf{rigorous evaluation on genuinely unseen identities}---with SECS in a standard speaker-verification space, McNemar significance, and a frozen-vs.\ soft-tuning ablation---showing that the face's contribution to vocal identity is real but modest, plus a working \textbf{zero-shot cross-lingual transfer} to Spanish.
\end{itemize}

The remainder of this paper is organized as follows. Section~\ref{sec:method} describes the proposed Freeze-Align framework. Section~\ref{sec:setup} details the experimental setup and evaluation protocol. Section~\ref{sec:results} presents and discusses the results. Finally, Section~\ref{sec:conclusions} concludes the paper.

%==================================================================================
% Section 3: Proposed Method
%==================================================================================
\section{Proposed Method}
\label{sec:method}

\subsection{Architectural overview}
\begin{figure}[t]
  \centering
  \includegraphics[width=\linewidth]{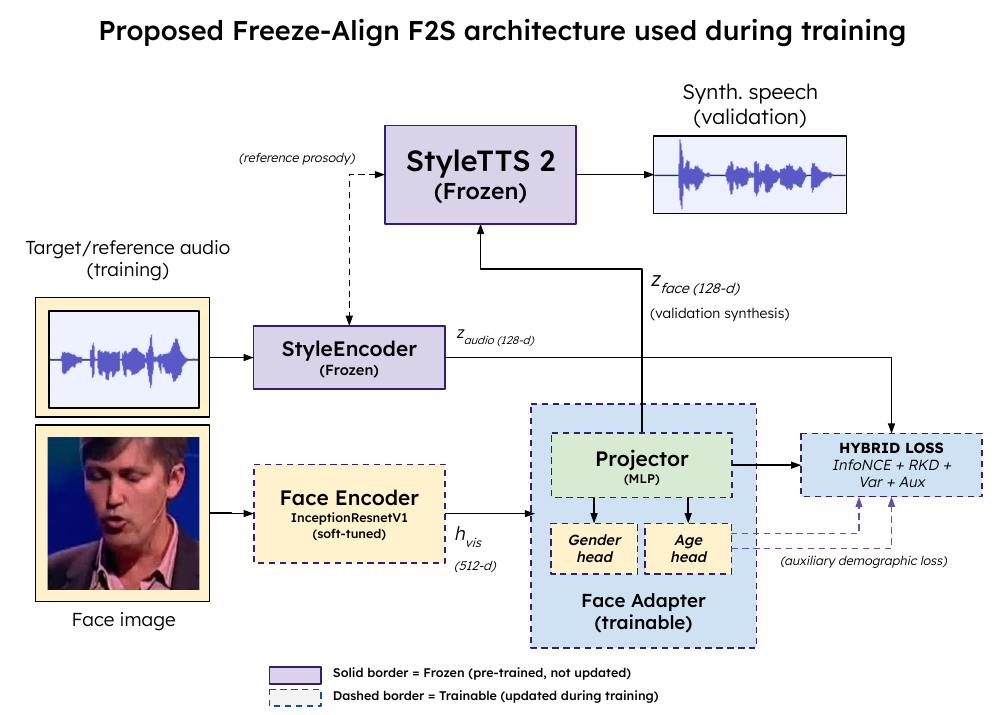}
  \caption{Proposed Freeze-Align F2S architecture used during training. An
  InceptionResnetV1 backbone maps the input face to a 512-d embedding $h_{vis}$, which the
  trainable Face Adapter (MLP, $512 \to 1024 \to 1024 \to 128$) projects into
  the 128-d style space $z_{face}$. The frozen StyleTTS 2 style encoder provides the
  acoustic target $z_{audio}$. Training aligns the two modalities through the hybrid
  loss (InfoNCE, RKD, variance, and demographic auxiliary heads). The acoustic teacher
  is frozen; the Face Adapter is trainable and the upper blocks of the face encoder are
  soft-tuned.}
  \label{fig:architecture}
\end{figure}

The overall architecture (Fig.~\ref{fig:architecture}) comprises a visual backbone, a trainable projection module (Face Adapter), and a frozen acoustic generator (StyleTTS 2 \cite{Li23-STTS2}). We choose StyleTTS 2 because its style-diffusion mechanism condenses timbre and prosody into a compact style vector decoupled from content---exactly the interface our adapter targets, enabling the decoupling of Section~\ref{ssec:decoupling}---while reaching human-level naturalness with modest public corpora (LibriTTS, $245$\,h \cite{Zen19-LibriTTS}) and open weights that make Spanish fine-tuning feasible.

Given a static face image $I$, an InceptionResnetV1 \cite{Schroff15-FaceNet} pre-trained on VGGFace2 \cite{Cao18-VGG} produces a 512-dimensional face embedding $h_{vis}$. As this embedding and the StyleTTS 2 style space differ in dimensionality and content (visual identity vs.\ acoustic timbre), we train a Face Adapter---an MLP ($512 \to 1024 \to 1024 \to 128$, $\approx$1.7M parameters, ReLU, $0.3$ dropout)---that projects $h_{vis}$ into the 128-dimensional style space, yielding $z_{face}$. We additionally \emph{soft-tune} the encoder's upper blocks ($\approx$18.3M parameters, most of the backbone) so the visual features specialize towards vocally-relevant traits, keeping the lower, generic layers frozen. The frozen StyleTTS 2 style encoder provides the acoustic target $z_{audio}$ (LibriTTS for English, fine-tuned on a Spanish corpus for the cross-lingual setting; see Section~\ref{sec:setup}).

% ----------------------------------------------------------------------------
% FIGURE 1: Architecture (single column).
% ----------------------------------------------------------------------------

\subsection{Hybrid loss formulation}
To avoid mode collapse and capture distinctive vocal characteristics, we use a multi-term loss:
\begin{equation}
\mathcal{L}_{total} = \lambda_{nce}\mathcal{L}_{NCE} + \lambda_{rkd}\mathcal{L}_{RKD} + \lambda_{var}\mathcal{L}_{Var} + \mathcal{L}_{Aux}
\label{eq:total_loss}
\end{equation}
This hybrid loss trains the Face Adapter and the soft-tuned face-encoder blocks; the StyleTTS 2 acoustic teacher remains frozen. The four terms play complementary roles: identity discrimination ($\mathcal{L}_{NCE}$), structure preservation ($\mathcal{L}_{RKD}$), collapse prevention ($\mathcal{L}_{Var}$), and demographic grounding ($\mathcal{L}_{Aux}$). These terms are detailed below.

\subsubsection{Supervised contrastive loss}
Rather than regressing each face embedding onto its target voice embedding with an MSE loss, which over-smooths predictions in weakly correlated multimodal spaces and encourages mode collapse, we align the two modalities with a contrastive, CLIP-style objective \cite{Oord18-CPC}. Each projected face $z_{face}^{i}$ is pulled towards the voice embeddings of the same speaker and pushed away from those of all other speakers in the batch. Since the balanced sampler places $K=4$ utterances per speaker per batch, each face has multiple positives $P(i)$, giving a supervised, multi-positive InfoNCE loss:
\begin{equation}
    \resizebox{0.9 \columnwidth}{!}{$\displaystyle
        \mathcal{L}_{NCE} = -\frac{1}{N}\sum_{i=1}^{N} \frac{1}{|P(i)|} \sum_{p \in P(i)} \log \frac{\exp(\text{sim}(z_{face}^{i}, z_{audio}^{p})/\tau)}{\sum_{j=1}^{N} \exp(\text{sim}(z_{face}^{i}, z_{audio}^{j})/\tau)}
    $}
    \label{eq:infonce}
\end{equation}
where $\text{sim}(\cdot)$ is the cosine similarity, $N$ the batch size, and $\tau$ a learnable, CLIP-style temperature (initialized at $0.04$ for the identity-priority configuration and $0.07$ otherwise). Contrasting each anchor against all in-batch voices yields richer gradients than pair- or triplet-based metric learning.

\subsubsection{Relational Knowledge Distillation (RKD)}
Beyond the absolute position of each projection, we preserve the \emph{pairwise geometry} of the acoustic space with the distance-wise term of RKD \cite{Park19-RKD}: within a batch $\mathcal{B}$, the distance between any two projected face embeddings should match the distance between the corresponding pair of audio embeddings, so that voices that are close (or far apart) in the style space map to faces that are equally close (or far apart):
\begin{equation}
    \resizebox{0.9\columnwidth}{!}{$\displaystyle
        \mathcal{L}_{RKD} = \frac{1}{|\mathcal{B}|^2}\sum_{(i,j) \in \mathcal{B}} L_{\delta} \left( \psi(z_{face}^{i}, z_{face}^{j}), \psi(z_{audio}^{i}, z_{audio}^{j}) \right)
    $}
    \label{eq:rkd_dist}
\end{equation}
where $\psi(a,b)$ is the Euclidean distance $\lVert a-b\rVert_2$ normalized by the mean pairwise distance in the batch (making the term scale-invariant), and $L_{\delta}$ is the Huber loss, robust to outliers. We use only this distance-preserving term, not the angular RKD variant. Transferring this relational geometry preserves the local structure of the acoustic manifold in the visual projection.

\subsubsection{Variance regularization loss \texorpdfstring{($\mathcal{L}_{Var}$)}{(L\_Var)}}
To explicitly discourage representation shrinkage, we align the per-batch standard deviation of the projected embeddings with that of the audio embeddings:
\begin{equation}
\mathcal{L}_{Var} = \frac{1}{D}\left\Vert \sigma(Z_{face}) - \sigma(Z_{audio}) \right\Vert_2^2
\label{eq:var_loss}
\end{equation}
where $Z_{face}$, $Z_{audio}$ stack the batch's $z_{face}$, $z_{audio}$ vectors, $\sigma(\cdot)$ is the per-dimension standard deviation across the batch and $D$ the embedding dimensionality (the audio standard deviation is treated as a fixed target).

\subsubsection{Demographic auxiliary loss \texorpdfstring{($\mathcal{L}_{Aux}$)}{(L\_Aux)}}
To retain essential biological markers, parallel heads predict gender and age, acting as semantic anchors during training:
\begin{equation}
\mathcal{L}_{Aux} = \lambda_{gen}\mathcal{L}_{BCE}(\hat{y}_{gen}, y_{gen}) + \lambda_{age}\mathcal{L}_{MSE}(\hat{y}_{age}, y_{age})
\label{eq:aux_loss}
\end{equation}

\subsection{Inference-time feature decoupling}
\label{ssec:decoupling}
A static image carries no temporal information about prosody, so conditioning solely on $z_{face}$ risks monotonous speech. We therefore decouple timbre from prosody at inference, exploiting that the StyleTTS 2 style vector splits into a timbre component (consumed by the waveform decoder) and a prosody component (consumed by the duration and prosody predictor). The face supplies the timbre, while each component can additionally be sampled from the native text-conditioned style diffusion model ($z_{diff}$) or taken from optional reference audio ($z_{ref}$). The two components are interpolated independently:
\begin{equation}
\begin{aligned}
z_{final}^{\,timbre} &= (1-\alpha)\, z_{face} + \alpha\, z_{diff}^{\,timbre},\\
z_{final}^{\,prosody} &= (1-\beta)\, z_{ref} + \beta\, z_{diff}^{\,prosody},
\end{aligned}
\label{eq:decoupling}
\end{equation}
where a low $\alpha$ preserves the face-derived identity (identity strength is thus $\propto 1-\alpha$) and $\beta$ sets how much prosodic expressiveness is drawn from the diffusion sampler; when no reference audio is provided, $z_{ref}$ defaults to $z_{face}$. Section~\ref{ssec:alpha} quantifies how $\alpha$ trades identity for naturalness, exposing this balance as an explicit control.

%==================================================================================
% Section 4: Experimental Setup
%==================================================================================
\section{Experimental Setup}
\label{sec:setup}

\subsection{Datasets and preprocessing}
We evaluate our proposed method in both English and Spanish. The Face Adapter is trained and evaluated on LRS3 \cite{Afouras18-LRS3}, a dataset of English TED-talk videos. Starting from the validated record list of prior F2S work and applying quality filtering (low-quality audio, undetectable faces, audiovisual desynchronization), we obtain $14{,}170$ aligned face--audio pairs spanning $2{,}007$ distinct speakers. Demographic labels (gender and age) for $1{,}449$ of them are obtained with InsightFace \texttt{buffalo\_l}\footnote{\url{https://github.com/deepinsight/insightface}}. Audio is resampled to $24$\,kHz and one random frame per clip serves as the face image (LRS3 clips are already head-cropped). For Spanish, since no comparable public audiovisual dataset exists, we do not train the adapter; instead, we only fine-tune the frozen StyleTTS 2 backbone on a private audio-only multi-speaker corpus ($48$ speakers, $21$M\,/\,$27$F, $53{,}336$ utterances) and apply the English-trained adapter zero-shot. Table~\ref{tab:datasets} summarizes both datasets.

\begin{table}[t]
  \caption{Datasets used in this work. LRS3 demographic statistics are
  automatically estimated with InsightFace over the labeled subset.}
  \label{tab:datasets}
  \centering
  \small
  \begin{tabular}{lcc}
    \toprule
     & \textbf{LRS3} & \textbf{Spanish corpus} \\
    \midrule
    Language            & English & Spanish \\
    Role                & adapter train/eval & backbone fine-tune \\
    Samples             & $14{,}170$ & $53{,}336$ \\
    Speakers            & $2{,}007$ & $48$ \\
    \ \ (with label)    & $1{,}449$ & --- \\
    Gender (M/F)        & $868/581$ & $21/27$ \\
    Age (mean$\pm$std)  & $41.0 \pm 11.1$ & $30 \pm 5$ \\
    \bottomrule
  \end{tabular}
  \vspace*{-0.5cm}
\end{table}

\subsection{Training configuration}
The adapter is trained on an NVIDIA A100 with PyTorch/\texttt{accelerate}, using AdamW \cite{Loshchilov19-AdamW} (lr $2\times10^{-3}$, cosine annealing with warm restarts) for $200$ epochs. The batch size is scaled to $1024$ via a custom balanced sampler that places $K{=}4$ samples per speaker in each batch, sharpening speaker boundaries. We train the Face Adapter together with a soft-tuning of the face encoder's upper blocks, for $\approx$20.0M trainable parameters in total: $\approx$1.7M in the adapter and $\approx$18.3M (about $78\%$ of the backbone) in the encoder; only the StyleTTS 2 acoustic teacher stays frozen. The reported checkpoint is selected on a disjoint development set by an objective criterion that combines identity and naturalness while penalizing collapse, rather than peak training retrieval.

\subsection{Evaluation protocol and metrics}
\label{ssec:metrics}
We evaluate on a disjoint test set of $24$ LRS3 speakers (from the pretrain partition, with any identity overlapping the training set removed), none seen during training. Over this pool we report cross-modal face$\to$voice retrieval: for each face we predict $z_{face}$ and rank the $N{=}24$ real-audio style embeddings by cosine similarity; Top-$k$ is the fraction of faces whose own speaker ranks among the $k$ most similar (chance Top-1 $\approx 4.17\%$, Top-5 $\approx 20.8\%$). We also measure identity fidelity at the embedding level as the mean cosine similarity between each face-predicted embedding $z_{face}^{i}$ and the audio style prototype $\bar{z}_{audio}^{\,s(i)}$ of its speaker (the mean of that speaker's real-audio style embeddings) as follows:
\begin{equation}
\text{SECS}_{\text{emb}} = \frac{1}{N} \sum_{i=1}^{N} \text{sim}\left(z_{face}^{i}, \bar{z}_{audio}^{\,s(i)}\right)
\label{eq:secs}
\end{equation}
We additionally report SECS$_{\text{audio}}$, the cosine similarity between speaker embeddings of the \emph{synthesized} and the real audio, computed with an external speaker-verification encoder (Resemblyzer, a GE2E model \cite{Wan18-GE2E}) to match the protocol of prior F2S work. Diversity is measured with SED (Speaker Embedding Diversity), the mean pairwise cosine similarity among all predicted embeddings, where a \emph{lower} value indicates more diverse (less collapsed) voices:
\begin{equation}
\text{SED} = \frac{2}{N(N-1)} \sum_{i<j} \text{sim}\left(z_{face}^{i}, z_{face}^{j}\right)
\label{eq:sed}
\end{equation}
Statistical significance of retrieval differences is assessed with McNemar's test.

\subsection{Model configurations}
\label{ssec:configs}
We compare four loss configurations, summarized in Table~\ref{tab:configs} ($\lambda_{rkd}{=}0.5$ throughout). Because they differ in $\lambda_{nce}$ and $\tau$ as well as in the demographic and variance terms, Table~\ref{tab:results_main} compares operating points rather than a single-factor ablation.

\begin{table}[t]
\centering
\caption{Loss configurations compared in this work. All use $\lambda_{rkd}{=}0.5$. A dash (--) indicates a disabled term.}
\label{tab:configs}
\begin{tabular}{lccccc}
\toprule
Configuration & $\lambda_{nce}$ & $\tau$ & $\lambda_{gen}$ & $\lambda_{age}$ & $\lambda_{var}$ \\
\midrule
Balanced          & $1.0$ & $0.07$ & $0.2$  & $0.1$  & --    \\
Identity-priority & $2.0$ & $0.04$ & $0.05$ & $0.01$ & --    \\
+\,variance       & $1.0$ & $0.07$ & $0.2$  & $0.1$  & $1.0$ \\
Pure-contrastive  & $1.0$ & $0.07$ & --     & --     & --    \\
\bottomrule
\end{tabular}
\vspace*{-0.5cm}
\end{table}
%==================================================================================
% Section 5: Results and Discussion
%==================================================================================
\section{Results and Discussion}
\label{sec:results}

\subsection{Comparison with prior F2S work}
For context, prior end-to-end F2S systems report SECS on synthesized audio of $0.754$ (FVTTS \cite{Lee24-FVTTS}) and $0.748$ (Face-TTS \cite{Lee23-ImaginaryVoice}) on LRS3, while our SECS$_{\text{audio}}$ is $0.61$ (EN) and $0.57$ (ES, cross-lingual). These are \emph{not directly comparable}: prior systems fine-tune the full TTS end-to-end, whereas we keep the acoustic teacher frozen and adapt only the face side, and our SECS$_{\text{audio}}$ partly reflects the StyleTTS~2 voice prior (Section~\ref{ssec:alpha}), so the gap is not a pure identity difference. A protocol-matched comparison (end-to-end fine-tuning of StyleTTS~2 on LRS3) is left as future work.

\subsection{Results on unseen identities}
\label{ssec:main}
Table~\ref{tab:results_main} reports retrieval and identity on the disjoint unseen set for the four configurations (English and Spanish) plus a frozen-encoder ablation. Face$\to$voice retrieval is consistently above chance (Top-1 around $2\times$, Top-5 $1.6$--$2.1\times$), indicating a weak but real biometric signal for speakers never seen in training; the predictions stay diverse (Table~\ref{tab:results_main}, SED below the real-audio level), so we see no mode collapse. With only $24$ unseen identities, Top-1 is noisy: a McNemar test finds \emph{no significant difference} among the four configurations (all $p>0.1$), so the hybrid objective is robust and no single term dominates, rather than one configuration being clearly best. Conditioned on Spanish text, the English-trained adapter transfers zero-shot to fluent Spanish (ES rows), with comparable retrieval and identity, so the face-to-style mapping is largely language-agnostic.

\begin{table}[t]
  \caption{Results on the disjoint set of $24$ unseen LRS3 identities (chance level: Top-1 $4.17\%$, Top-5 $20.8\%$). SECS$_{\text{emb}}$ is the embedding-level cosine; lower SED is more diverse. EN and ES embedding metrics are computed in each language's own style space and are not cross-comparable.}
  \label{tab:results_main}
  \centering
  \begin{tabular}{lcccc}
    \toprule
    \textbf{Config} & \textbf{Top-1} & \textbf{Top-5} & \textbf{SECS$_{\text{emb}}$} & \textbf{SED} \\
    \midrule
    \multicolumn{5}{l}{\textit{English}} \\
    Balanced          & 9.3\%  & 35.1\% & 0.40 & 0.36 \\
    Identity-priority & 8.9\%  & 32.6\% & 0.42 & 0.43 \\
    \ + variance      & 7.1\%  & 34.0\% & 0.41 & 0.37 \\
    Pure-contrastive  & 9.3\%  & 39.2\% & 0.41 & 0.39 \\
    \ \ \emph{Frozen enc.} (abl.) & 6.8\% & 38.8\% & 0.29 & 0.23 \\
    \midrule
    \multicolumn{5}{l}{\textit{Spanish (cross-lingual)}} \\
    Balanced          & 12.7\% & 44.0\% & 0.55 & 0.50 \\
    Identity-priority & 7.1\%  & 35.5\% & 0.53 & 0.52 \\
    \ + variance      & 12.7\% & 41.1\% & 0.42 & 0.32 \\
    Pure-contrastive  & 10.2\% & 38.0\% & 0.52 & 0.50 \\
    \bottomrule
  \end{tabular}
\end{table}

\noindent\textbf{Adaptation depth (frozen vs.\ soft-tuning).}
Soft-tuning the upper blocks is decisive for absolute alignment: it nearly doubles the embedding-level fidelity (SECS$_{\text{emb}}$ $0.42$ vs.\ $0.29$ for a frozen encoder), though the two are indistinguishable on Top-1 ($p{=}0.25$). A frozen encoder ranks reasonably but places the style off the acoustic manifold, degrading synthesis; we adapt only the upper blocks because, with $\approx$2k identities, full fine-tuning risks catastrophic forgetting.

\subsection{Perceptual quality (automatic proxies)}
\label{ssec:perceptual}
Lacking a large-scale listening test, we report automatic proxies for subjective quality: UTMOS \cite{Saeki22-UTMOS} (naturalness) and NISQA \cite{Mittag21-NISQA} (signal quality), plus SECS$_{\text{audio}}$ (identity). Both MOS predictors are English-biased, so Spanish values are only indicative. Table~\ref{tab:results_perceptual} reports means over synthesized utterances of unseen identities (inference at $\alpha{=}0.5$); the reported checkpoint epoch was selected on this unseen development split, a mild optimism we declare. Ground-truth recordings are an upper anchor.

\begin{table}[t]
  \caption{Audio identity and quality on unseen identities ($\alpha{=}0.5$). SECS$_{\text{audio}}$ via an external speaker encoder; UTMOS/NISQA as MOS proxies. Ground truth applies to English recordings.}
  \label{tab:results_perceptual}
  \centering
  \begin{tabular}{lccc}
    \toprule
    \textbf{System} & \textbf{SECS$_{\text{audio}}$} & \textbf{UTMOS} & \textbf{NISQA} \\
    \midrule
    \multicolumn{4}{l}{\textit{English}} \\
    Ground truth      & ---  & 3.61 & 3.63 \\
    Balanced          & 0.62 & 3.72 & 3.94 \\
    Identity-priority & 0.62 & 3.72 & 3.84 \\
    \ + variance      & 0.61 & 3.98 & 4.22 \\
    Pure-contrastive  & 0.61 & 3.95 & 4.21 \\
    \midrule
    \multicolumn{4}{l}{\textit{Spanish (cross-lingual)}} \\
    Balanced          & 0.57 & 2.73 & 3.66 \\
    Identity-priority & 0.57 & 2.79 & 3.74 \\
    \ + variance      & 0.57 & 2.67 & 3.74 \\
    Pure-contrastive  & 0.57 & 2.88 & 3.85 \\
    \bottomrule
  \end{tabular}
\end{table}

In English the synthesized voices reach UTMOS $3.7$--$4.0$, on par with or above ground truth ($3.61$), confirming that conditioning the frozen backbone on the predicted style does not degrade naturalness, even for held-out (unseen) identities. The cross-lingual Spanish values are lower (UTMOS $2.7$--$2.9$): both proxies are English-biased and the Spanish backbone is reached zero-shot, so these are indicative rather than conclusive.

\subsection{Identity--naturalness control ($\alpha$)}
\label{ssec:alpha}
The decoupling weight $\alpha$ of Eq.~\eqref{eq:decoupling} exposes an explicit identity-vs-naturalness trade-off (Table~\ref{tab:alpha}): identity (SECS$_{\text{audio}}$) peaks at $\alpha{=}0.3$---a near-pure face vector lands off-manifold and hurts naturalness---while higher $\alpha$ blends towards the diffusion prior, trading identity for naturalness. We use $\alpha{=}0.5$ as a balanced point.\footnote{The $\alpha$ sweep is a separate evaluation run (fixed prompts, $\beta{=}0.9$); its $\alpha{=}0.5$ point matches Table~\ref{tab:results_perceptual} only approximately ($\lesssim 0.1$ UTMOS).} SECS$_{\text{audio}}$ ranges only from $0.56$ (diffusion-dominant) to $0.64$ (face-dominant), so the face supplies a modest part of the absolute identity, most coming from the StyleTTS~2 prior; absent a mismatched-face baseline, this range bounds the face's net contribution.

\begin{table}[t]
  \caption{Effect of the decoupling weight $\alpha$ (English, unseen identities, $\beta{=}0.9$).}
  \label{tab:alpha}
  \centering
  \begin{tabular}{lccccc}
    \toprule
    $\alpha$ & 0.1 & 0.3 & 0.5 & 0.7 & 0.9 \\
    \midrule
    SECS$_{\text{audio}}$ & 0.61 & \textbf{0.64} & 0.62 & 0.57 & 0.56 \\
    UTMOS                 & 1.82 & 2.66 & 3.62 & \textbf{4.17} & 4.10 \\
    \bottomrule
  \end{tabular}
\end{table}

%==================================================================================
% Section 6: Conclusions
%==================================================================================
\section{Conclusions}
\label{sec:conclusions}
We presented a Freeze-Align framework that synthesizes a plausible, coherent voice from a single face by adapting a face-recognition embedding to the style space of a frozen StyleTTS 2 teacher, via a trainable Face Adapter and soft-tuning of the face encoder. The mapping holds for unseen identities, with naturalness on par with ground truth (UTMOS $3.7$--$4.0$ vs.\ $3.61$), above-chance retrieval, and zero-shot transfer to fluent Spanish. An inference-time control shows the face contributes a modest but real part of the identity, most coming from the TTS prior. Future work includes an end-to-end protocol-matched comparison, a human MOS/AB study, native multilingual training, and stochastic adapters.
%==================================================================================
% Section 7: Acknowledgments and Generative AI Disclosure
%==================================================================================
\section{Acknowledgments}
\ifcameraready
    This work was supported by Monoceros Labs; the R\&D\&I project C-HUM-223-UGR23, co-financed by the Consejer\'{\i}a de Universidad, Investigaci\'on e Innovaci\'on and the European Union through the FEDER Andalusia 2021--2027 Programme; and grants PID2022-141378OB-C22 and AIA2025-163317-C32, funded by MICIU/AEI/10.13039/501100011033 and ERDF/EU.
\else
    Acknowledgments withheld for double-blind review.
\fi

\section{Generative AI Use Disclosure}
The authors used generative AI tools for language editing and coding assistance only, not to generate scientific content, design the methodology, or interpret the results; the authors reviewed all content and take full responsibility for the publication.

%==================================================================================
% References (IEEE format via BibTeX)
%==================================================================================
\bibliographystyle{IEEEtran}
\bibliography{mybib}

\end{document}